# Analysis of Local Anisotropy Fluctuations in Compact Objects

Manuel Malaver[1] y María Esculpi[2]

[1] Maritime University of the Caribbean, Department of Basic Sciences, Catia La Mar, Venezuela
 E-mail: mmf.umc@gmail.com

[2] Central University of Venezuela, Department of Applied Physics, Faculty of Engineering, Caracas, Venezuela.
 E-mail: mariaesculpiw@gmail.com

**Abstract**

 Mathematical modeling within the framework of the general theory of relativity has been used to explain the behavior and structure of massive objects as neutron stars, quasars, black holes, pulsars and white dwarfs and requires finding the exact solutions of the Einstein-Maxwell system. In this paper we study the effects induced by fluctuations of local anisotropy in a new family of anisotropic solutions depending on a parameter α, whose value α=2 provides a radial pressure having the same functional dependence on the radial coordinate as the Schwarzschild solution. It is shown the effect the functional dependence on the radial coordinate has in the occurrence of cracking within the sphere when anisotropy fluctuations are allowed.
**Keywords:** fluctuations, local anisotropy, cracking, radial pressure, Schwarzschild solution

## 1. Introduction

   The study of the relation between compacts objects and the gravitational collapse is one of the most fundamental and important factors in astrophysics and has attracted much researchers and scientists due to formulation of the general theory of relativity.  In the construction of the first theoretical models of relativistic stars, some works are important such as Schwarzschild [1], Tolman [2], Oppenheimer and Volkoff [13]. Schwarzschild [1] found exact solutions to the Einstein's Field Equations and Tolman [2] proposed a method in order to obtain explicit solutions of static spheres of fluid in terms of known analytical functions. Oppenheimer and Volkoff [3] have deployed Tolman's solutions in order to investigate about gravitational balance of neutron stars. It is noticed that Chandrasekhar's contributions [4] in modelling for production of white dwarfs under relativistic effects and research of Baade and Zwicky [5] establish the concept of neutron stars as relativistic star of very dense matter.

  Many researchers have used a great variety of mathematical techniques to try in order to obtain solutions of the Einstein-Maxwell field equations since it has been demonstrated by Bowers y Liang [6], Ruderman [7], Canuto [8], Komathiraj and Maharaj [9], Cosenza *et al* [10],  Esculpi *et al* [11] and Malaver [12-22].  These investigations show that the system of Einstein-Maxwell equations plays an important role to describe ultracompacts objects.

  In the formulism of realistic model of super dense stars, it is also important to include the pressure anisotropy. Bowers and Liang [4] extensively discuss the effect of pressure anisotropy in general relativity. At a density of the order of $10^{15}$ $g/cm^3$ nuclear matter may be anisotropic when its interactions need to be treated relativistically [7].

In massive objects the radial pressure may differ from the tangential. In 1933, Lemaitre [23] established that in stellar models consisting of spherically symmetric distribution of matter the stress tensor may be locally anisotropic. From theoretical work [24-33] realistic stellar models it has been suggested that superdense matter may be anisotropic, at least in some density ranges. The existence of anisotropy within a star can be explained by the presence of a solid core, phase transitions, a type III super fluid [34], a pion condensation [35] or another physical phenomenon by the presence of an electrical field [36].

Bowers and Liang [4] generalized the equation of hydrostatic equilibrium for the case of local anisotropy. Bondi [31] have shown that for anisotropic fluids there exist a surface redshift bound, if either the strong or dominant energy condition are considered to hold within the star. Esculpi et al [11] have obtained a new family of anisotropic solutions with uniform energy density, where solutions depend on two parameters that can be adjusted to improve gravitational redshift. Also the assumption of local anisotropy has been used to study problems related to various relativistic compact objects [37-47].

In order to describe the behavior of an anisotropic fluid distribution when it exits the dynamic equilibrium, Herrera [48] and Di Prisco et al. [49,50] propose the concept of cracking, which implies the appearance of different radial forces within the system. We say that there are cracking whenever inward in the inner part of the sphere for all values of the radial coordinate. Otherwise, when the force is directed outward inside and changes direction in the outermost regions of the star, then there is an inversion [48-50].

Herrera [48] established that the appearance of a cracking is induced by the local anisotropy of a fluid distribution, whereas in the case of a perfect fluid outside equilibrium, the configuration tends to expand or collapse. Chan et al [51] studied the role of local anisotropy over dynamic instability and found that small anisotropies can drastically change the evolution of a system. Di Prisco et al [50] studied the role of local anisotropy fluctuations and determined that these fluctuations are a crucial factor for cracking. Abreu et al [41] considered a particular type of perturbation in which the difference in sound velocities is taken into account, where $v^2_{sr}$ and $v^2_{s\perp}$ represent the velocity of radial and tangential sound, respectively and found that regions where $v^2_{sr} < v^2_{s\perp}$, within a matter distribution, no cracking will occur and it could be considered as stable. Manjarrés [52] analyzes what happens in charged spheres when the charge is perturbed with energy density and anisotropy, and finds that these perturbations can lead to the appearance of cracking. Malaver [53] finds that when a slow adiabatic contraction is performed on an anisotropic sphere model, which depends on an adjustable parameter and a coefficient that measures anisotropy, this model turns out to be unstable on the surface of the sphere so instability could occur in the outer layers.

The aim of this research is to study the behavior of anisotropic fluid solutions found by Esculpi et al [11] against density variations and local anisotropy and to determine the factors that favor the appearance of cracking. The results shall be compared with previous results for similar anisotropic solutions. We have used the method suggested by Herrera [48] and Di Prisco et al [50] in the study of cracking for compact and anisotropic objects with constant density in which radial pressure is only a function of the radial coordinate. In our case radial pressure is written as the product of a function that depends on the anisotropy factor and a function of the radial coordinate, which as in Herrera's work [48] remains constant against the variation of energy density and anisotropy. Comparing the behavior of the family of solutions for different values of the parameter α, which defines the functional form of the pressure with the radial coordinate, the influence of the model on the appearance of cracking is verified. The paper is structured as follows: the next section, Sect.2, are presented the interior solutions of Einstein-Maxwell field equations of anisotropic fluid. In Sect.3, the occurrence of cracking was calculated when local anisotropy fluctuations occurred for an anisotropic star model with uniform energy density. In Sect.4 discusses and concludes the work.

## 2. The Einstein Field Equations

Considering a spherically symmetrical quadridimensional space, whose line element is described by the Schwarzschild coordinates [1,2] given by:

$$ds^2 = e^\nu dt^2 - e^\lambda dr^2 - r^2(d\theta^2 + \sin^2\theta d\phi^2) \tag{1}$$

With a static distribution of matter consisting of a non-pascalian hydrodynamic fluid, with an energy-impulse tensor given by the expression

$$T^{\mu\nu} = (\rho + P_t)U^\mu U^\nu - P_t g^{\mu\nu} + (P_r - P_t)\chi^\mu \chi^\nu \tag{2}$$

Einstein's field equations are:

$$8\pi T^0_{\ 0} = 1/r^2 - e^{-\lambda}(1/r^2 - \lambda'/r) \tag{3}$$

$$8\pi T^1_{\ 1} = 1/r^2 - e^{-\lambda}(1/r^2 + \nu'/r) \tag{4}$$

$$8\pi T^2_{\ 2} = 8\pi T^3_{\ 3} = \frac{-1}{4}e^{-\lambda}(2\nu'' + \nu'^2 - \lambda'\nu' + 2(\nu' - \lambda')/r \tag{5}$$

and from equation (3) we obtain

$$e^{-\lambda} = 1 - 2m/r \quad \text{where} \quad m(r) = \int 4\pi \rho r^2 dr \tag{6}$$

Using equations (4) and (5) we obtain the generalized Tolman-Oppenheimer-Volkov equation [3] for hydrostatic equilibrium in the presence of tangential pressure:

$$\frac{dP_r}{dr} = -(\rho + P_r)\frac{4\pi P_r r^3 + m(r)}{r^2(1-2m(r)/r)} + 2\frac{(P_t - P_r)}{r} \tag{7}$$

## 3. Analysis of fluctuations in anisotropic stars

This section determines the appearance of cracking when fluctuations in local anisotropy occur for an anisotropic star model with uniform energy density proposed by Esculpi et al [11]. The procedure suggested by Herrera [48] and Di Prisco et al [49,50] has been used in the study of cracking for compact and anisotropic objects.

There are a large number of physical processes that give rise to deviations from the local isotropy of the fluid, such as exotic phase transitions involving the appearance of an anisotropic phase during the gravitational collapse process [51]. The existence of solid nuclei and the presence of superfluids may give rise to local anisotropy [34]. Also, the overlap of two perfect fluids can be described as an anisotropic fluid [54].

The Eq. (7) can be written as

$$R = \frac{dP_r}{dr} + \frac{4\pi r P_r^2}{1-2m/r} + \frac{P_r m}{r^2(1-2m/r)} + \frac{4\pi r \rho P_r}{1-2m/r} + \frac{\rho m}{r^2(1-2m/r)} - \frac{2(P_t - P_r)}{r} \tag{8}$$

Where $R$ defines the total radial force on each fluid element. If the system under study is taken out of equilibrium by some perturbation, a total radial force R appears, which can lead to cracking or inversions [48,49]. Given a state equation for hydrodynamic variables and suitable juncture conditions, it is possible to obtain a solution for Einstein's field equations. A distribution of matter with uniform energy density $\rho$, contained in a sphere of radius a, is considered and a relationship between radial and tangential pressures is proposed which generalizes the solution proposed by Dev and Gleiser [35] as follows:

$$\rho = \rho_0 \quad r \leq a \tag{9}$$

$$P_t - P_r = \frac{2\pi C r^2(P_r^2 + \alpha P_r \rho + \rho^2)}{3(1 - \frac{2m(r)}{r})} \tag{10}$$

Where $C$ is the anisotropy factor and $\alpha$ is a new parameter that varies the relationship between radial and tangential pressure for a given value of the anisotropy parameter. Equation (10) can be substituted into the equation for hydrostatic equilibrium and solved, considering the possible values for the discriminant and obtain:

$$\Delta = \rho_0^2\left[(4-\alpha C)^2 - 4(3-C)(1-C)\right] \tag{11}$$

For values $\Delta > 0$, the radial pressure within the star is given by:

$$P_r = \rho_0 \left[ \frac{1-C}{\beta+\Gamma} \right] \left[ \frac{(1-2m/r)^{\Gamma/2} - (1-2M/R)^{\Gamma/2}}{(1-2M/R)^{\Gamma/2} - \left(\frac{\beta-\Gamma}{\beta+\Gamma}\right)(1-2m/r)^{\Gamma/2}} \right] \qquad (12)$$

For the analysis of cracking, the solution shown in equation (12) representing a new exact solution for anisotropic stars with uniform density where $C$ is the anisotropy constant, $\beta = 2\left(1 - \frac{\alpha C}{4}\right)$, $\Gamma = \left[\beta^2 - (3-C)(1-C)\right]^{1/2}$ and α is a parameter that measures the degree of anisotropy. For α=2 an expression is obtained for the radial pressure which has the same functional dependence on the radial coordinate of the Schwarzschild solution. The following dimensionless variables are now introduced:

$$\mu = 1 - 2M/a \quad y \quad x = r/a \qquad (13)$$

The expression (12) can then be written in the form:

$$P_r = \rho_0 f(c)\varphi(x)$$

(14)

and we have :

$$\varphi(x) = \frac{\left[1 - (1-\mu)x^2\right]^{\Gamma/2} - \mu^{\Gamma/2}}{\mu^{\Gamma/2} - \left[\frac{\beta-\Gamma}{\beta+\Gamma}\right]\left[1 - (1-\mu)x^2\right]^{\Gamma/2}}$$

(15)

$$f(c) = \frac{1-C}{\beta+\Gamma}$$

(16)

The system is now perturbed according to the scheme established by Herrera [49] and Di Prisco et al [50,51] where density and anisotropy are also perturbed and the radial dependence is invariant, i.e.:

$$\tilde{P}_r = \tilde{\rho}_0 \tilde{f}(C)\varphi(x) \qquad (17)$$
$$\tilde{C} = C + \delta C \qquad (18)$$

$$\tilde{\rho}_0 = \rho_0 + \delta\rho_0 \tag{19}$$

where it has been considered that:

$$\gamma = \tilde{\rho}_0 / \rho_0 \tag{20}$$

The tilde indicates how much is being disturbed.

From equations (12), (13), (14) and (15) the expression for R takes the form:

$$R = \rho_0 \frac{f(C)}{a} \frac{d\varphi(x)}{dx} + \frac{x}{a} \frac{(1-\mu)\rho_0}{2} \frac{\left[f^2\varphi(x)^2(3-C)+(1-C)+f(C)\varphi(x)(4-\alpha C)\right]}{\left[1-(1-\mu)\alpha x^2\right]} \tag{21}$$

To calculate $\tilde{R}$, the following dimensionless function must be introduced:

$$\tilde{\hat{R}} = a\tilde{R}/\rho_0 \tag{22}$$

and the expression for $\tilde{\hat{R}}$ is as:

$$\tilde{\hat{R}} = \tilde{f}(C)\frac{d\varphi(x)}{dx} + \gamma^2 x \frac{(1-\mu)}{2} \frac{\left[f^2\varphi^2(3-C)+(1-C)+f\varphi(x)(4-\alpha C)\right]}{\left[1-(1-\mu)\gamma x^2\right]} \tag{23}$$

Considering

$$\delta\tilde{\hat{R}} = \frac{\partial \tilde{\hat{R}}}{\partial \gamma}\delta\gamma \, |_{\tilde{C}=C}^{\gamma=1} + \frac{\partial \tilde{\hat{R}}}{\partial C}\delta C \, |_{\tilde{C}=C}^{\gamma=1} \tag{24}$$

We obtain that :

$$\delta\tilde{\hat{R}} = \left\{ f\frac{d\varphi(x)}{dx} + \frac{x(1-\mu)}{\left[1-(1-\mu)x^2\right]^2}\left[1-\frac{(1-\mu)x^2}{2}\right]\left[1-C+f^2\varphi^2(3-C)+(4-\alpha C)f\varphi(x)\right] \right\}\delta\gamma$$

$$+ \left( \frac{\partial f}{\partial C}\frac{d\varphi(x)}{dx} + \frac{x(1-\mu)}{2\left[1-(1-\mu)x^2\right]}\left\{ -1-f^2\varphi^2-\alpha f\varphi(x)+\frac{\partial f}{\partial C}\left[2f(3-C)\varphi^2+\varphi(x)(4-\alpha C)\right] \right\} \right)\delta C$$

$$\tag{25}$$

For a cracking to occur it is necessary that $\tilde{\hat{R}}$ has a zero in the range $-1 \leq x \leq 1$.

Figure 1 shows how the radial force varies with the radius of the star for an anisotropy factor value C=0.73 and different values of $\alpha$, keeping the gravitational potential value fixed and equal to $\mu = 0.2$ and where it has been considered that $P_r \geq 0, \Delta > 0$. It is shown that when $\alpha$ increases the radial force $\tilde{\hat{R}}$ decreases; for values of $\alpha <2$ sign changes occur and the cracking occurs in regions close to the surface of the star as it increases $\alpha$, which corresponds to the presence of cracking for these values of $\alpha$. In this model, as in that of Bowers and Liang [6], cracking is presented for a low value of $\mu$ [51], that is for more compact configurations. An analogous behavior is presented in Figure 2 for an anisotropy factor of $C = 0.45$ and the same value of the gravitational potential. In both figures it is observed as an increase of C causes a decrease in the radial force $\tilde{\hat{R}}$, contrary to what occurs in the model of Bowers and Liang [6], in which it is observed that as it decreases h, where h=1-2C, which is the parameter that measures anisotropy, increases the radial force, as shown in Figure 3. For the models considered, small fluctuations in the values of C and h, that is, changes in the local anisotropy of the fluid can cause the appearance of cracking.

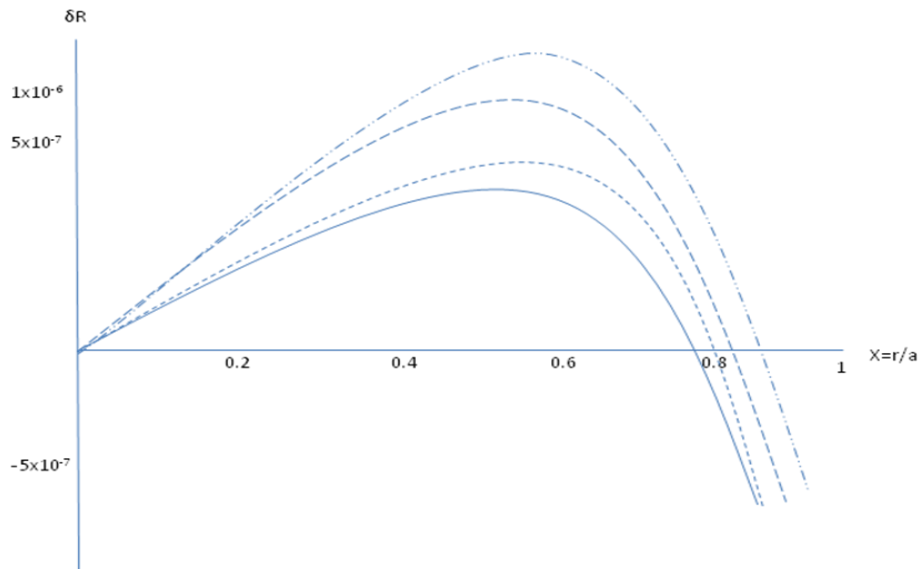

**Figure 1.** $\delta\hat{R}$ as a function of x for $\mu$=0.2, $C$=0.73 and different values of the parameter $\alpha$. The plot with lines and with two alternating dots corresponds to $\alpha = 0.25$, the dash line corresponds to $\alpha$=0.5, with short lines corresponds to $\alpha$=1.0 and solid line is for $\alpha$=1.5.

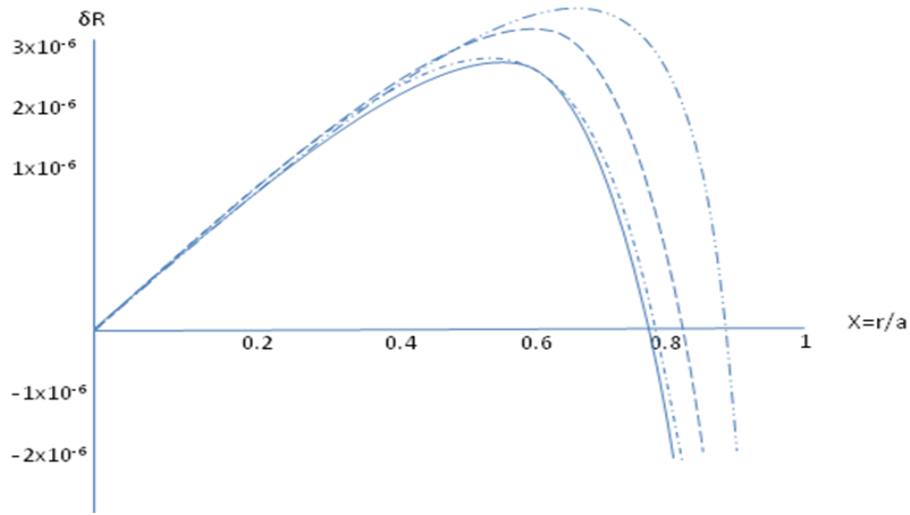

**Figure 2.** $\delta\hat{R}$ as a function of x for $\mu=0.2$, $C=0.45$ and different values of the parameter $\alpha$. The plot with lines and with two alternating dots corresponds to $\alpha = 0.25$, the dash line corresponds to $\alpha=0.5$, with short lines corresponds to $\alpha=1.0$ and solid line is for $\alpha=1.5$.

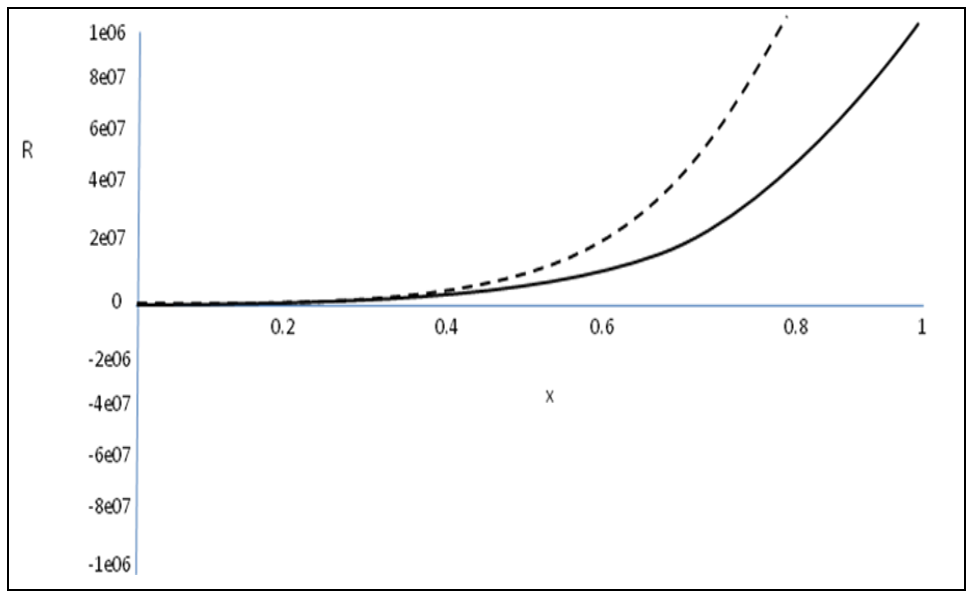

**Figure 3.** $\delta\hat{R}$ as a function of $x$ for $\mu=0.2$ for the Bowers and Liang model [6]. The solid line and the long-dash line correspond to $C=0.45$; $h=0.1$ and $C=0.73$; $h = -0.46$, respectively.

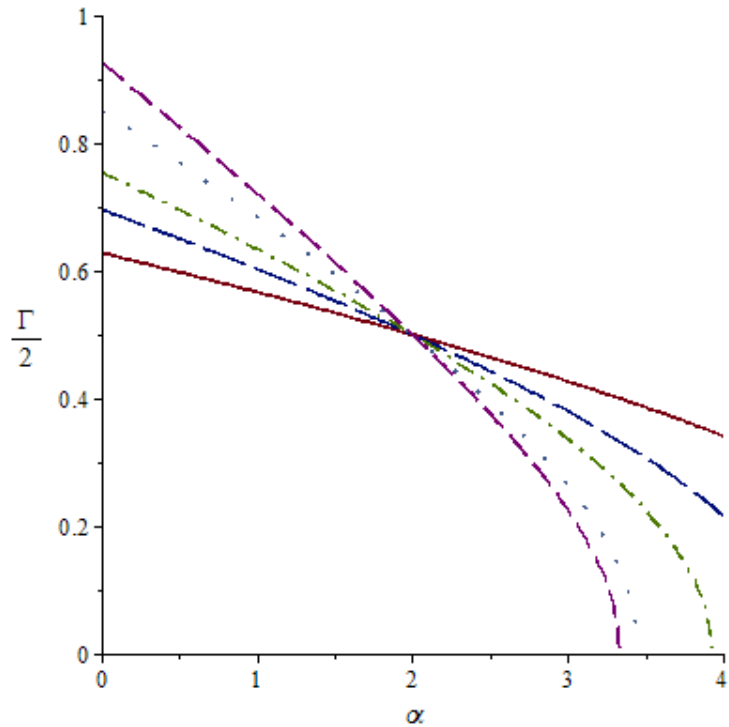

**Figure 4.** Exponent $\Gamma/2$ as a function of the parameter $\alpha$ for different values of the anisotropy factor C. The solid lines, long dash line, dash-dot line, spaced dots and short-dash line correspond to the values of C = 0.15, 0.25, 0.35, 0.55 and 0.75 respectively.

## 4. Conclusions

A cracking analysis for a new anisotropic star model with uniform energy density has been presented in this paper. For this model with anisotropy parameter values $C=0.45$ and $C=0.73$, cracking occurs near the surface of the sphere as values of $\alpha$ increase and anisotropy is an important parameter for determining cracking conditions when gravitational potential is modified.

It is interesting to highlight the marked dependence of the response to the cracking with the type of model. Variations in anisotropy are expected to allow for cracking under certain conditions. A modification of the parameter $\alpha$ can generate different expressions for radial pressure, which in turn changes the response to the cracking. As each value of $\alpha$ changes the exponent $\Gamma$ it is obvious then that the parameter $\alpha$ defines the different model types, as shown in Figure 4. It is observed that $\alpha$ is increased when $\Gamma$ decreases and takes values smaller than one for values of $\alpha$ greater than 2, and greater than one for $\alpha < 2$. For $\alpha = 2$, the exponent acquires the constant value $\Gamma = 1$, regardless of the value of C. If $\alpha > 2$, the exponent $\Gamma/2$ decreases when the value of C increases, keeping constant the value of $\alpha$. Si $\alpha < 2$, $\Gamma/2$ decreases when C decreases for a fixed value of $\alpha$. For $\alpha < 2$ the presence of cracking is observed.

The appearance of cracking, associated with fluctuations of the local anisotropy, depends on the functional form of the pressure with the radial coordinate. Indeed, a modification of the parameter can generate different expressions for radial pressure, which in turn changes the response to the cracking, but such a cracking always occurs in regions close to the surface of the sphere. This behavior is to be expected according to Malaver [53] who finds that when a slow adiabatic contraction is made in the model of anisotropic star with uniform density proposed by Esculpi et al [11], this model turns out to be more stable in the outer layers than that of the Bowers and Liang solution [6], so it is likely to present instability in the outer layers.